# Optical gain in colloidal quantum dots is limited by biexciton absorption, not biexciton recombination

Davide Zenatti[1] and Patanjali Kambhampati[1]*

[1]Department of Chemistry, McGill University, Montreal, QC, H3A 0B8, Canada

*pat.kambhampati@mcgill.ca

**Abstract**

Despite three decades of experimental study, optical gain in colloidal quantum dots still lacks a microscopic theory capable of explaining gain thresholds approaching one exciton per dot, their size dependence, or the anomalously small effective stimulated-emission cross sections observed across materials. Existing descriptions treat quantum dots as effective two-level systems comprised of an exciton and a biexciton, attributing gain thresholds to biexciton Auger recombination. This assumption is inconsistent with state-resolved optical pumping experiments and basic spectroscopic constraints. Here we present a microscopic theory of optical gain explicitly anchored in the Einstein relations governing absorption and stimulated emission. Within this framework, gain is determined by a spectral balance between stimulated emission from single excitons and excited-state absorption into biexcitonic manifolds, rather than by biexciton lifetimes. Using a spin–boson description of excitons coupled to a lattice bath, we show that gain thresholds and effective gain cross sections are controlled by the interplay of biexciton stabilization and exciton–lattice dressing. The theory unifies disparate materials by quantitatively explaining all longstanding gain phenomenology in CdSe quantum dots and predicts a continuous crossover to effective four-level, near-thresholdless gain in dynamically disordered lattices such as perovskite quantum dots.

Optical gain underpins the technological impact of colloidal quantum dots in light-emitting devices, lasers, and active photonic materials[1-10]. Despite almost three decades of experimental study across a wide range of compositions and architectures, the microscopic origin of optical gain in quantum dots remains unresolved[7, 11-14]. In particular, there is no theory that explains why gain thresholds asymptotically approach one exciton per dot, why their size dependence is weak or inverted[8, 13, 15], or why effective stimulated-emission cross sections are consistently far smaller than expected[7, 13, 15-26]. The persistence of these discrepancies points not to experimental limitations, but to a failure of the theoretical framework used to interpret optical gain physics in colloidal quantum dots.

Nearly all descriptions of colloidal quantum-dot gain adopt an effective two-level picture in which gain thresholds are assumed to be set by biexciton Auger recombination[1-7, 9-10]. Within this phenomenological view, gain is achieved when biexcitons are populated faster than they decay, and improvements in gain performance are therefore attributed to Auger suppression [1-5, 21-23, 25, 27-28].

However, this framework is inconsistent with well-established spectroscopic observations. It neglects the intrinsic Stokes shift[29-33] between absorption and emission, fails to account for state-resolved measurements[13, 15] showing weak size dependence of gain thresholds, and offers no microscopic connection between gain thresholds, effective gain cross sections, and photoluminescence quantum yield. As a result, it provides no predictive basis for understanding gain across different classes of quantum dots, from crystalline covalent systems such as CdSe to dynamically disordered ionic systems such as lead-halide perovskites[16, 34-37].

A physically consistent theory of optical gain must instead begin from constraints that hold universally. Chief among these are the Einstein relations[11, 14], which link absorption and stimulated emission and thereby fix the spectral structure of gain. In much of the quantum-dot gain literature, these constraints are not enforced: stimulated emission is inferred phenomenologically from photoluminescence, Stokes shifting is treated ad hoc, and absorption into biexcitonic manifolds is neglected or treated as a secondary correction [1-5, 23]. Yet biexcitonic absorption competes directly with stimulated emission in the same spectral window [38-43] and therefore plays a central role in determining gain thresholds and effective gain cross sections.

Here we present a microscopic theory of optical gain in colloidal quantum dots explicitly anchored by the Einstein relations. Using a spin–boson description of excitons coupled to a lattice bath, we treat stimulated emission, Stokes shifting, phonon or polaron dressing, and excited-state absorption into biexcitonic manifolds on equal footing. Within this framework, gain thresholds and effective gain cross sections emerge as spectral quantities governed by the energetic placement of biexcitonic absorption relative to emission and by the strength of exciton–lattice coupling, rather than by multiexciton lifetimes. The theory quantitatively explains longstanding gain phenomenology in CdSe quantum dots and predicts a continuous crossover to effective four-level, near-thresholdless gain in dynamically disordered lattices. Together, these results establish a unified and predictive framework for optical gain in colloidal quantum materials.

**Figure 1** defines the microscopic degrees of freedom that control optical gain in colloidal quantum dots. An exciton is not a single optical transition, but a manifold of fine-structure–split states arising from exchange, spin–orbit coupling, and confinement anisotropy[29, 32-33, 44-48]. These splittings exist independently of lattice relaxation and generate an intrinsic Stokes shift between absorption and emission. Optical excitation of this manifold therefore immediately opens access to higher-lying many-body states.

In particular, exciton population necessarily enables transitions into the biexciton sector[13, 15, 40, 42-43, 49], which is itself a dense manifold of correlated two-electron–two-hole configurations rather than a single level at twice the exciton energy. Absorption from the single-exciton manifold into these biexcitonic states constitutes excited-state absorption (ESA), an intrinsic optical loss channel that spectrally overlaps with stimulated emission and competes directly with gain. Despite this inevitability, biexcitonic absorption is typically neglected or treated phenomenologically in gain models.

The projection of this electronic structure into optical spectra is controlled by coupling to the lattice bath. In crystalline covalent quantum dots such as CdSe, exciton–lattice coupling is relatively weak (with Huang-Rhys parameters of S ~ 0.1) and dominated by narrow phonon modes[50], producing modest reorganization energies and strong spectral correlation between absorption and emission. In contrast, dynamically disordered ionic lattices, such as lead-halide perovskite quantum dots, exhibit strong coupling (S~1) to a broadband phonon bath, resulting in substantial spectral dressing in the form of phonon or polaronic envelopes[51-53].

As illustrated schematically in **Fig. 1b,** lattice coupling governs not only the excitonic Stokes shift but also the energetic placement and spectral breadth of biexcitonic absorption relative to emission. The resulting overlap between stimulated emission and ESA—and therefore the balance between gain and loss—is set jointly by excitonic fine structure and system–bath coupling. **Figure 1** thus establishes the central premise of this work: optical gain in quantum dots is governed by the combined electronic–lattice landscape, rather than by exciton or biexciton lifetimes.

**Figure 2** shows directly why effective two-level descriptions of optical gain fail in quantum dots. Optical pumping does not simply bleach the ground-state exciton absorption and generate stimulated emission from a single transition. Because excitons are composite many-body states with internal structure, excitation inevitably opens optical pathways into biexcitonic manifolds [38-43]. As illustrated in **Fig. 2a,** exciton absorption and emission coexist spectrally with biexciton absorption and emission channels [54] that are offset but not well separated from the excitonic transitions[13, 15]. The main point spectrally is that stimulated emission from an exciton will arise at a Stokes shifted energy of the spontaneous PL band, and loss into biexcitons is between absorption and emission involving single excitons. Hence the problem of gain and loss is fundamentally a spectral problem not a lifetime problem.

State-resolved optical pumping measurements[13, 15] make the consequences explicit. Even when the band-edge exciton absorption is nearly fully bleached, the nonlinear response remains only weakly negative. The small magnitude of the stimulated-emission feature, despite strong bleaching, demonstrates that optical gain is not governed by population inversion of a single excitonic transition, but by the coexistence of competing optical processes that remain active under excitation.

This competition is revealed by decomposing the nonlinear response into gain and loss channels **(Fig. 2b).** Stimulated emission contributes a negative signal, while excited-state absorption (ESA) from the exciton into biexcitonic states contributes a positive signal of comparable magnitude [38-43]. These contributions are spectrally displaced yet substantially overlapping across the emission band, producing strong cancellation in the net response. As a result, the effective gain cross section is determined not by the intrinsic strength of stimulated emission, but by the degree of spectral overlap between emission and biexcitonic absorption.

This overlap is an intrinsic property of the exciton–biexciton energy landscape and cannot be removed by increasing population inversion or extending biexciton lifetimes. **Figure 2** therefore isolates the central microscopic problem of optical gain in quantum dots: gain and loss are spectrally entangled at the many-body level, and optimizing gain requires controlling the relative energetic placement of excitonic emission and biexcitonic absorption.

**Figure 3** recasts optical gain in quantum dots within the familiar language of laser level schemes{Siegman, 1986 #6217} while making explicit why the canonical two-level models fail without modification. The left panel shows the commonly assumed two-level system (2LS), consisting of a ground state $|G\rangle$ and a single exciton $|X\rangle$. This picture is frequently invoked[1-3, 5, 23, 25]—often implicitly—in discussions of quantum-dot lasing, where gain is presumed to occur once the average exciton number exceeds unity and to be limited primarily by biexciton Auger recombination.

For quantum dots, however, the 2LS is internally inconsistent. Exciton population immediately opens optical transitions into biexcitonic states [38-43] $|XX\rangle$, which lie below $2E_X$ by the biexciton binding energy and therefore introduce absorptive pathways that compete directly with stimulated emission. Moreover, a true two-level system cannot support gain under incoherent pumping[55], underscoring the incompatibility of the 2LS picture with the conditions under which quantum-dot gain is observed.

The center panel introduces the minimal physically sensible extension: a three-level system (3LS) arising from excitonic fine structure[44]. The exciton manifold splits into an absorptive state $|X_A\rangle$ and a lower-energy emissive state $|X_E\rangle$, producing an intrinsic Stokes shift even in the absence of strong lattice reorganization. This structure captures a central experimental reality—that absorption and emission originate from distinct electronic states—and is essential for interpreting state-resolved gain spectroscopy. However, the 3LS does not remove loss: excitation from $|X_E\rangle$ into biexcitonic states such as $|X_E X_A\rangle$ produces excited-state absorption (ESA) that spectrally overlaps with stimulated emission. In this regime, gain is controlled by the energetic placement and strength of these biexcitonic absorption channels, not by biexciton lifetimes.

The right panel shows the emergence of an effective four-level system (4LS) generated by stronger coupling to the lattice bath. Phonon dressing or polaron formation displaces absorption and emission surfaces along nuclear coordinates, reducing spectral overlap between emissive exciton states and absorptive biexcitonic manifolds. This lattice-dressed 4LS suppresses reabsorption and enables optical gain at substantially lower excitation densities.

As shown below, crystalline covalent quantum dots such as CdSe remain close to the 3LS limit, whereas dynamically disordered lattices, including lead-halide perovskites, approach the effective 4LS regime. **Figure 3** thus establishes the conceptual framework for the remainder of the paper: optical gain in quantum dots is governed not by population counting or Auger timescales, but by the interplay between excitonic fine structure, biexcitonic absorption, and system–bath coupling.

Any physically consistent theory of optical gain must begin from the Einstein relations, which encode microscopic reversibility and detailed balance[55]. The detailed derivation is in the Supplementary Information (**SI**). For an optical transition between two many-body states $|i\rangle$ and $|f\rangle$, the absorption and stimulated-emission cross sections are not independent but share the same intrinsic spectral lineshape. Explicitly, for a transition of frequency $\omega_{fi}$,

$$\sigma_{\text{SE}}(\omega) = \frac{g_f}{g_i}\,\sigma_{\text{ABS}}(\omega), \qquad (1)$$

where $g_i$ and $g_f$ are the degeneracies of the initial and final states. Equation (1) is exact and independent of material, dimensionality, or disorder. Its immediate consequence is that optical gain cannot be constructed by assigning an arbitrary stimulated-emission spectrum; any suppression of gain must arise from additional absorption channels that cancel stimulated emission within the same spectral window.

In quantum dots, these channels follow directly from the minimal electronic structure. A quantum dot consists not of a single excitonic transition, but of an excitonic fine-structure manifold [44-47] and a dense manifold of biexcitonic states [38-43]. The electronic Hamiltonian may be written schematically as

$$H_{\text{el}} = E_G\,|G\rangle\langle G| + \sum_{\alpha} E_{X_\alpha}\,|X_\alpha\rangle\langle X_\alpha| + \sum_{\beta} E_{XX_\beta}\,|XX_\beta\rangle\langle XX_\beta|. \qquad (2)$$

Here $|X_\alpha\rangle$ denotes fine-structure exciton states and $|XX_\beta\rangle$ denotes biexciton configurations stabilized by Coulomb correlations. Optical pumping of the exciton manifold therefore necessarily opens transitions $|X_\alpha\rangle \to |XX_\beta\rangle$, giving rise to excited-state absorption (ESA) that competes directly with stimulated emission. Whereas most optical gain experiments in colloidal QD are performed with non-resonant excitation at 3.1 eV [1-5, 23], greater precision is obtained with state-resolved optical pumping [13, 15, 56-59] performed resonantly with the band edge exciton, or higher-lying excitons to probe gain bandwidth control.

These electronic manifolds are embedded in a lattice environment that cannot be treated perturbatively. Excitons and biexcitons couple diagonally to lattice displacements, which we describe using a spin–boson Hamiltonian,

$$H = H_{\text{el}} + \sum_k \hbar \omega_k b_k^\dagger b_k + \sum_n |n\rangle\langle n| \sum_k g_{nk}(b_k^\dagger + b_k), \qquad (3)$$

where $n \in \{X_\alpha, XX_\beta\}$. This coupling displaces nuclear equilibrium configurations upon electronic excitation, producing both fine-structure–induced Stokes shifts and phonon or polaron progressions depending on the bath spectral density.

Optical spectra follow from dipole–dipole correlation functions of this open quantum system. Within the cumulant (independent-boson) formalism, the absorption or emission lineshape associated with a given transition is

$$\sigma(\omega) \propto \text{Re} \int_0^\infty dt\ e^{i(\omega-\omega_0)t - g(t)}, \qquad (4)$$

where the lineshape function $g(t)$ is determined entirely by the phonon spectral density $J(\Omega)$. Absorption and emission therefore retain the same intrinsic spectral structure required by Eq. (1), but are displaced in energy by lattice relaxation.

Optical gain is governed by the difference between two Einstein-consistent spectra: stimulated emission from excitons and excited-state absorption into biexcitonic manifolds. The experimentally observed gain cross section is

$$\sigma_{\text{eff}}(\omega) = \sigma_{\text{SE}}(\omega) - \sigma_{\text{ESA}}^{XX}(\omega), \qquad (5)$$

where $\sigma_{\text{ESA}}^{XX}$ arises from transitions $|X_\alpha\rangle \to |XX_\beta\rangle$. Because both contributions scale with exciton population, gain is achieved only when their spectral overlap is sufficiently reduced.

Two dimensionless parameters control this balance. The first characterizes the energetic placement of biexcitonic absorption relative to exciton emission,

$$\chi \equiv \frac{\Delta_{XX}^A}{\Delta_{SS}}, \qquad (6)$$

where $\Delta_{XX}^A$ is the mean exciton–biexciton absorption offset and $\Delta_{SS}$ is the exciton Stokes shift. The second parameter characterizes exciton–lattice coupling,

$$\eta \equiv \frac{\Delta_{\text{FS}}}{\Delta_{\text{ph}}}, \qquad (7)$$

where $\Delta_{\text{FS}}$ is the exciton fine-structure splitting and $\Delta_{\text{ph}}$ is the width of the phonon or polaron progression. Large $\eta$ corresponds to crystalline, fine-structure–dominated systems such as CdSe, while small $\eta$ corresponds to glassy or polaronic lattices such as lead-halide perovskites.

The gain threshold follows from the spectral condition

$$\int d\omega\, \sigma_{\text{eff}}(\omega) = 0, \qquad (8)$$

yielding an average exciton threshold $\langle N_{\text{th}}\rangle$ determined entirely by $\chi$ and $\eta$. Notably, biexciton lifetimes do not enter Eq. (8): Auger recombination limits aspects of gain dynamics, not gain onset. In the limit $\chi \approx 1$ and large $\eta$, stimulated emission and ESA remain strongly correlated and cancel, pinning $\langle N_{\text{th}} \rangle \approx 1$, as observed in CdSe quantum dots[13, 15]. Reducing $\chi$ and $\eta$ progressively decorrelates loss from emission, driving a continuous transition to effective four-level, near-thresholdless gain.

With the microscopic framework established, we benchmark the theory against the most precise experimental phenomenology available for optical gain in CdSe quantum dots (**Fig. 4**). State-resolved optical pumping measurements isolate intrinsic gain physics from ensemble and cavity effects and therefore provide a stringent test. **Figure 4** addresses three observations that have resisted a unified explanation: the dependence of gain thresholds on photoluminescence quantum yield (PLQY), the origin of the small effective stimulated-emission cross section, and the weak size dependence of the gain threshold $\langle N_{\text{th}} \rangle$.

**Figure 4a** shows the evolution of the gain threshold with surface passivation for a fixed dot size[13, 15, 56-57]. While improved passivation is known empirically to increase PLQY, its connection to gain thresholds has not been formulated microscopically. In the present framework, PLQY reflects the partitioning of exciton population between band-edge states and interfacial traps. Trapping introduces additional excited-state absorption channels into biexcitonic manifolds involving surface-localized excitons, which are more strongly stabilized than band-edge biexcitons and therefore shift absorptive transitions deeper into the emission band. The resulting increase in spectral overlap between stimulated emission and biexcitonic absorption raises $\langle N_{\text{th}} \rangle$. As passivation improves and PLQY increases, these loss channels are suppressed and the gain threshold decreases, without altering intrinsic oscillator strengths or violating Einstein relations.

The same mechanism governs the effective stimulated-emission cross section (**Fig. 4b**). In this framework, the small measured $\sigma_{eff}$ in CdSe quantum dots arises from spectral cancellation between stimulated emission and excited-state absorption. As trapping-induced biexcitonic absorption is reduced with increasing PLQY, this cancellation weakens and the effective gain cross section increases. The observed behavior follows directly from the Einstein-consistent balance of gain and loss and resolves the long-standing discrepancy between large absorption cross sections and small effective stimulated-emission cross sections.

**Figure 4c** addresses the size dependence of optical-gain thresholds in CdSe core quantum dots. State-resolved measurements show that $\langle N_{th} \rangle$ depends only weakly on dot diameter and is slightly lower for smaller (bluer) dots, in direct contradiction to Auger-based expectations. The theory reproduces this behavior and shows that it arises from spectral considerations: the gain threshold is set by the relative placement of excitonic emission and biexcitonic absorption, which depends on the size scaling of the Stokes shift and biexciton stabilization. A minimal model incorporating only the size dependence of the Stokes shift captures the correct qualitative trend, while the full theory yields quantitative agreement. By contrast, the commonly assumed 1/V scaling associated with Auger recombination[1-2, 4-5] fails both qualitatively and quantitatively.

Having established that the microscopic theory reproduces the gain phenomenology of CdSe quantum dots, we now show that the same framework becomes predictive when expressed in terms of two physically transparent control parameters. **Figure 5** presents a unified phase diagram for optical gain in quantum dots spanned by the dimensionless parameters $\eta$ and $\chi$. The parameter $\eta$ quantifies exciton–lattice coupling, ranging from narrow, crystalline phonon spectra to broad, glassy or polaronic baths, while $\chi \equiv \frac{\Delta_{XX}^A}{\Delta_{SS}}$ measures the energetic proximity of absorptive biexcitonic manifolds to the emission band and thus directly encodes biexciton-mediated loss. Both parameters correspond to concrete materials properties—biexciton stabilization, fine structure, and lattice dynamics—and are therefore accessible to synthetic control.

**Figures 5a** and **5b** summarize the behavior of the gain threshold ⟨N$_{th}$⟩ across this parameter space. In the crystalline limit ($\eta \approx 10$), representative of covalent CdSe quantum dots, absorptive biexciton transitions remain spectrally correlated with excitonic emission. This correlation enforces strong cancellation between stimulated emission and excited-state absorption, pinning ⟨N$_{th}$⟩ near unity over a broad range of $\chi$. The experimentally determined CdSe anchor point falls directly on this curve without adjustable parameters. As $\eta$ decreases and lattice dressing becomes stronger, biexcitonic absorption decorrelates from emission, driving a continuous crossover away from three-level–system behavior toward an effective four-level gain medium in which biexciton-mediated loss is progressively suppressed and gain thresholds are substantially reduced. Importantly, this reduction occurs even when biexciton stabilization remains finite, demonstrating that lattice coupling alone can drive near-thresholdless gain.

The same phase diagram simultaneously explains the long-standing puzzle of suppressed effective stimulated-emission cross sections in quantum dots, shown in **Figs. 5c and 5d.** While the bare stimulated-emission cross section is fixed by absorption through the Einstein relations, the experimentally measured cross section reflects the net balance between stimulated emission and excited-state absorption into biexcitonic manifolds. In crystalline CdSe quantum dots ($\eta \approx 10$), biexcitonic absorption remains spectrally locked to emission, producing strong cancellation and a reduced effective cross section despite a large Einstein-limited value. As lattice dressing increases ($\eta \ll 1$), absorptive biexciton transitions are spectrally displaced from emission, suppressing cancellation and allowing the effective cross section to approach its Einstein-limited value, as expected for an effective four-level gain medium.

Taken together, **Fig. 5** defines a predictive and unified design map for optical gain in colloidal quantum dots. Optimal gain performance is achieved not by engineering faster biexciton decay, but by minimizing biexciton-mediated spectral loss through controlled biexciton stabilization ($\chi$) while simultaneously exploiting lattice disorder or polaron formation ($\eta$) to decorrelate absorption from emission. This framework unifies decades of CdSe gain phenomenology with the exceptional gain properties of perovskite quantum dots and provides a concrete roadmap for realizing thresholdless optical gain in colloidal quantum materials.

Together, these results show that the defining characteristics of optical gain in CdSe quantum dots—thresholds exceeding one exciton, weak size dependence, and reduced effective gain cross sections—are governed by biexciton-mediated spectral loss controlled by surface chemistry and lattice coupling. Having established that decades of CdSe gain phenomenology follow from a single Einstein-consistent framework, we next use the same parameters to identify how these limits can be systematically overcome.

We have presented a microscopic theory of optical gain in colloidal quantum dots explicitly anchored by the Einstein relations, in which gain emerges as a spectral balance between stimulated emission and excited-state absorption into biexcitonic manifolds rather than as a competition between excitation and multiexciton decay. This framework resolves long-standing discrepancies in quantum-dot gain phenomenology—including thresholds exceeding one exciton, weak size dependence, and suppressed effective stimulated-emission cross sections—without invoking Auger-limited gain onset or ad hoc level schemes. By expressing gain behavior in terms of two physically transparent control parameters that quantify biexciton stabilization and exciton–lattice coupling, we establish a predictive phase diagram that unifies crystalline covalent and dynamically disordered ionic quantum dots and identifies a continuous route to effective four-level, near-thresholdless gain through lattice dressing. More broadly, these results reframe optical gain in quantum dots as an emergent property of coupled electronic and lattice degrees of freedom and provide a concrete foundation for the rational design of quantum-dot gain media and related light–matter systems.

**Acknowledgements**

P.K. acknowledges financial support from CFI, NSERC, and McGill University.

**Author contributions**

P.K. supervised the research. Theory and calculations and figures were done by DZ and PK. PK and DZ wrote the manuscript.

**Methods.**

Details of theory and calculations are in the Supplementary Information.

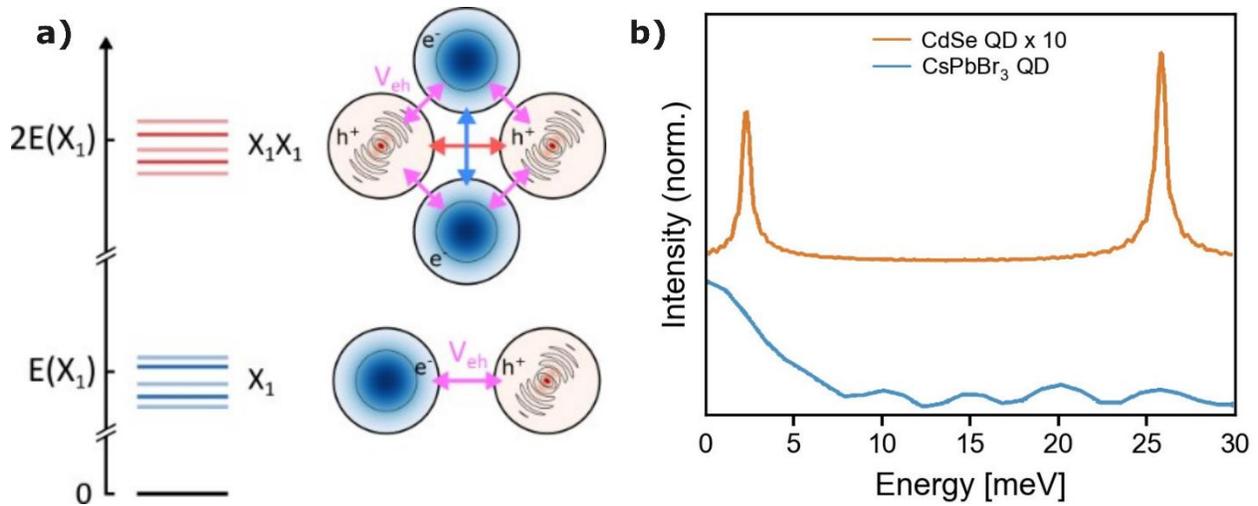

**Figure 1 | Microscopic ingredients governing optical gain in quantum dots. (a)** Electronic structure relevant to optical gain. The quantum-dot exciton comprises a fine-structure manifold of states $X_i$, whose splittings produce an intrinsic Stokes shift between absorption and emission. Optical excitation of the exciton manifold necessarily enables transitions into a dense biexciton manifold $X_i X_j$, giving rise to excited-state absorption that competes with stimulated emission. **(b)** Role of lattice coupling. Representative Raman spectra illustrate two limiting regimes: crystalline CdSe quantum dots with narrow phonon spectra and weak lattice reorganization, and lead-halide perovskite quantum dots with broadband phonon coupling and strong spectral dressing. Lattice coupling controls the energetic placement and spectral overlap of biexcitonic absorption and excitonic emission, thereby setting gain thresholds and effective gain cross sections. Data adapted from [53]

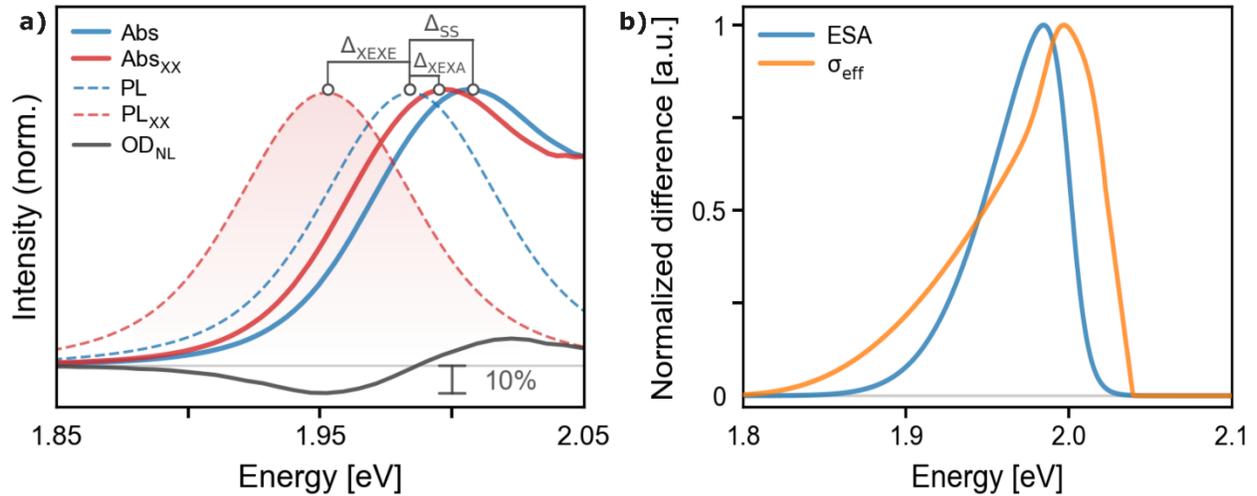

**Figure 2 | Spectroscopic origin of gain and loss in quantum dots. (a)** Representative spectra governing optical gain in CdSe quantum dots. In addition to exciton absorption ($Abs_X$) and photoluminescence ($PL_X$), optical pumping accesses biexcitonic transitions, including absorption into the biexciton manifold ($Abs_{XX}$) and biexciton emission ($PL_{XX}$). Shown is the nonlinear absorption spectrum under state-resolved optical pumping, where exciton absorption is nearly fully bleached. The remaining weak negative signal corresponds to stimulated emission, revealing strong cancellation by excited-state absorption. **(b)** Decomposition of the nonlinear response into gain and loss channels. Stimulated emission contributes a negative signal, while excited-state absorption into biexcitonic states contributes a positive signal. Their substantial spectral overlap produces strong cancellation across the emission band, resulting in a reduced effective gain cross section. Data adapted from state-resolved optical pumping measurements [13, 15].

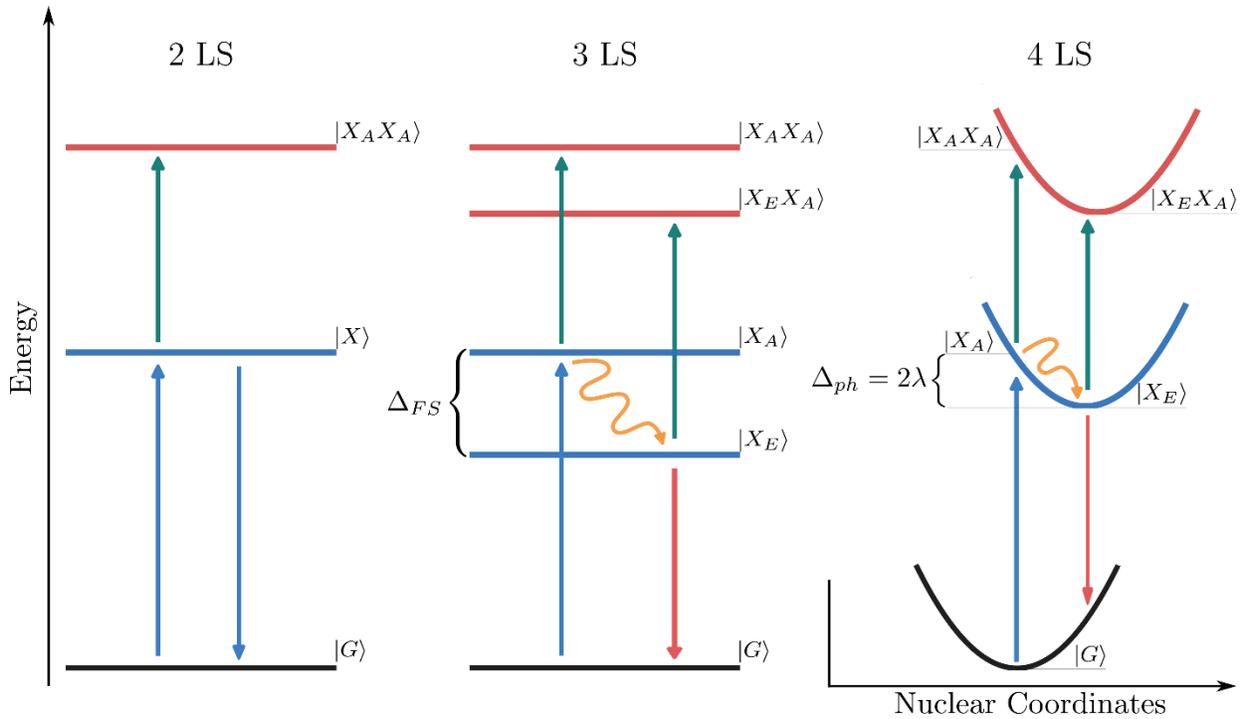

**Figure 3 | From few-level laser models to a microscopic description of optical gain in quantum dots.** Schematic evolution of laser level schemes adapted to colloidal quantum dots, highlighting the roles of exciton fine structure, biexcitonic states, and lattice coupling. **Left (2LS):** Two-level description $|G\rangle \leftrightarrow |X\rangle$ commonly assumed in quantum-dot gain models. This picture is internally inconsistent, as exciton population necessarily enables absorption into biexcitonic states $|XX\rangle$ and steady-state gain cannot be sustained under incoherent pumping. **Middle (3LS):** Incorporation of exciton fine structure splits absorption and emission into distinct states $|X_A\rangle$ and $|X_E\rangle$, producing an intrinsic Stokes shift and the minimal physically sensible gain model for quantum dots. In this regime, optical gain competes directly with excited-state absorption into biexcitonic manifolds such as $|X_E X_A\rangle$. **Right (4LS):** Strong coupling to the lattice bath displaces absorption and emission surfaces along nuclear coordinates, reducing spectral overlap between stimulated emission and biexcitonic absorption. This effective four-level system suppresses reabsorption and enables near-thresholdless gain.

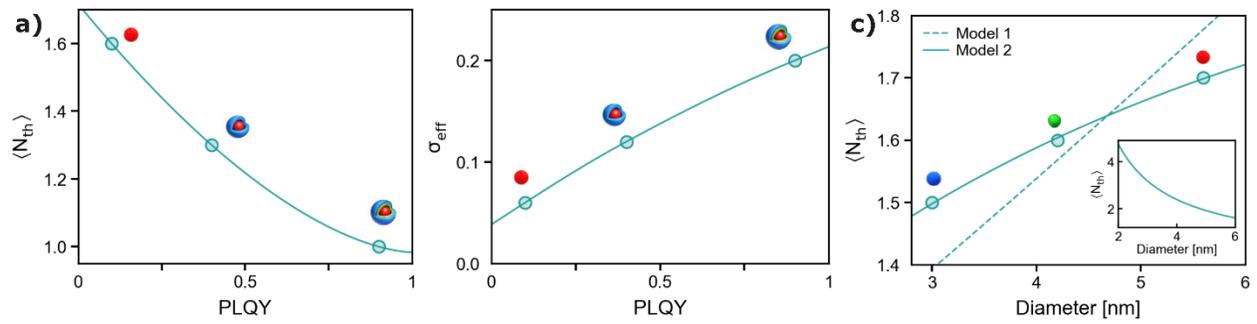

**Figure 4 | Microscopic theory reproduces state-resolved optical-gain phenomenology in CdSe quantum dots.** Comparison of theory with high-precision state-resolved optical pumping measurements [13, 15, 56-57] **(a)** Gain threshold ⟨$N_{th}$⟩ versus photoluminescence quantum yield (PLQY) for fixed dot size. The theory reproduces the observed reduction in ⟨$N_{th}$⟩ with increasing PLQY, arising from suppression of trapping-induced excited-state absorption into biexcitonic manifolds. **(b)** Effective stimulated-emission cross section versus PLQY. Reduced biexciton-mediated absorption increases the measured gain cross section, consistent with Einstein relations when all absorptive channels are included. **(c)** Size dependence of gain thresholds in CdSe core quantum dots. The weak, slightly inverted size dependence observed experimentally is reproduced by the theory and shown to arise from spectral placement of excitonic emission and biexcitonic absorption. A minimal Stokes-shift model captures the qualitative trend, while the full theory yields quantitative agreement; the commonly assumed 1/V Auger scaling (inset) fails.

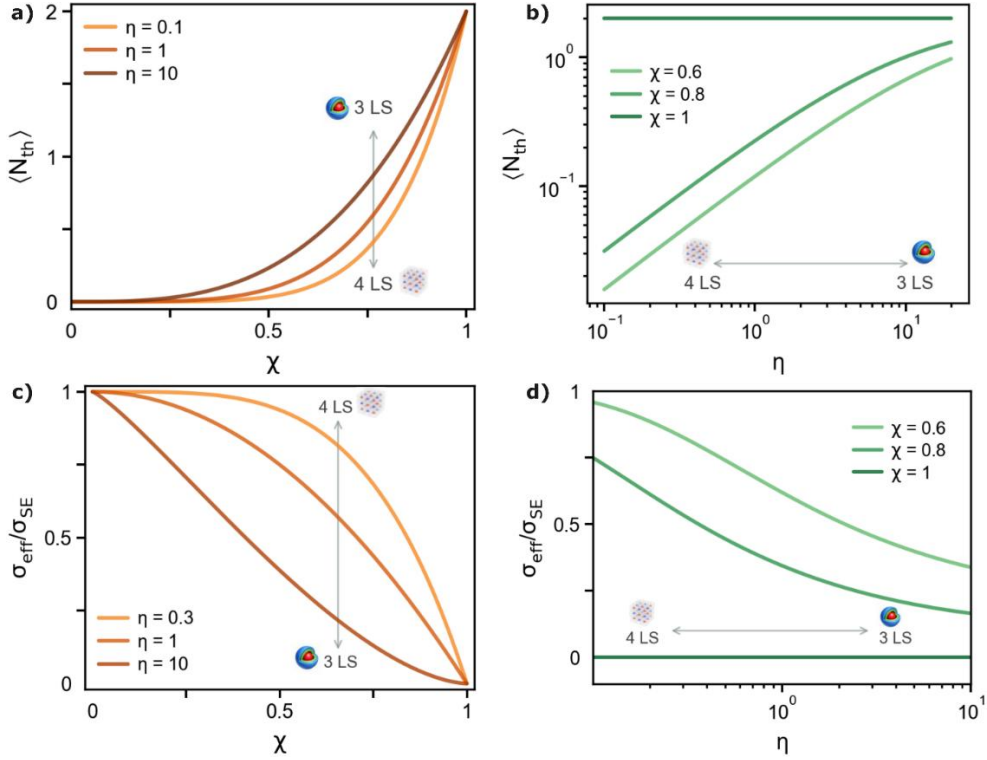

**Figure 5 | Predictive phase diagram for optical gain thresholds and effective gain cross sections in quantum dots.** Unified description of optical-gain thresholds and effective stimulated-emission cross sections in terms of two dimensionless control parameters: the biexciton proximity parameter $\chi$, which quantifies the energetic overlap of absorptive biexcitonic states with the emission band, and the lattice-coupling parameter $\eta$, which characterizes the strength and bandwidth of exciton–lattice dressing. **(a)** Calculated gain threshold $\langle N_{th} \rangle$ as a function of $\chi$ for fixed $\eta$. In the crystalline limit ($\eta \sim 10$), representative of CdSe quantum dots, absorptive biexciton states remain spectrally correlated with emission, enforcing strong cancellation and pinning $\langle N_{th} \rangle$ near unity (3LS regime). Stronger lattice dressing ($\eta \ll 1$) decorrelates absorption from emission, driving a continuous crossover to effective four-level gain with reduced thresholds (4LS regime). **(b)** $\langle N_{th} \rangle$ versus $\eta$ for fixed $\chi$, showing monotonic threshold reduction with increasing lattice disorder even at finite biexciton stabilization. **(c)** Effective stimulated-emission cross section $\frac{\sigma_{\text{eff}}}{\sigma_{\text{SE}}}$ as a function of $\chi$ for fixed $\eta$. Spectral overlap with biexcitonic absorption strongly suppresses $\sigma_{\text{eff}}$ in the crystalline (3LS) limit, while lattice dressing progressively restores the Einstein-limited value in the 4LS regime. **(d)** $\sigma_{\text{eff}}/\sigma_{\text{SE}}$ versus $\eta$ for fixed $\chi$, demonstrating monotonic enhancement of the effective gain cross section as biexcitonic absorption is spectrally displaced from emission. Filled markers indicate the CdSe anchor point derived from state-resolved optical pumping experiments. Arrows illustrate the materials-design trajectory from crystalline, three-level–like quantum dots to dynamically disordered or polaronically dressed systems exhibiting four-level, near-thresholdless optical gain.

References

1. Ahn, N.; Livache, C.; Pinchetti, V.; Klimov, V. I. Colloidal Semiconductor Nanocrystal Lasers and Laser Diodes. *Chemical Reviews* **2023,** *123* (13), 8251-8296
2. Park, Y.-S.; Roh, J.; Diroll, B. T.; Schaller, R. D.; Klimov, V. I. Colloidal quantum dot lasers. *Nature Reviews Materials* **2021,** *6* (5), 382-401
3. Jung, H.; Ahn, N.; Klimov, V. I. Prospects and challenges of colloidal quantum dot laser diodes. *Nature Photonics* **2021,** *15* (9), 643-655
4. García de Arquer, F. P.; Talapin, D. V.; Klimov, V. I.; Arakawa, Y.; Bayer, M.; Sargent, E. H. Semiconductor quantum dots: Technological progress and future challenges. *Science* **2021,** *373* (6555), eaaz8541
5. Pietryga, J. M.; Park, Y. S.; Lim, J.; Fidler, A. F.; Bae, W. K.; Brovelli, S.; Klimov, V. I. Spectroscopic and Device Aspects of Nanocrystal Quantum Dots. *Chem. Rev.* **2016,** *116*, 10513
6. Geiregat, P.; Van Thourhout, D.; Hens, Z. A bright future for colloidal quantum dot lasers. *NPG Asia Materials* **2019,** *11* (1), 41
7. Zenatti, D.; Kambhampati, P. A Path towards Thresholdless Colloidal Quantum Dot Lasers by Solving Decades of Mythology on Optical Gain. *arXiv preprint arXiv:2510.01199* **2025**,
8. Kambhampati, P. Quantum dots are beginning to lase in the blue. *Nature Nanotechnology* **2025,** *20* (2), 189-189
9. Wang, Y.; Sun, H. Advances and Prospects of Lasers Developed from Colloidal Semiconductor Nanostructures. *Prog. Quantum Electron.* **2018,** *60*, 1
10. Wang, Y.; Li, X.; Nalla, V.; Zeng, H.; Sun, H. Solution-processed low threshold vertical cavity surface emitting lasers from all-inorganic perovskite nanocrystals. *Advanced Functional Materials* **2017,** *27* (13), 1605088
11. Ryu, J.; Park, S. D.; Baranov, D.; Rreza, I.; Owen, J. S.; Jonas, D. M. Relations between absorption, emission, and excited state chemical potentials from nanocrystal 2D spectra. *Science Advances* **2021,** *7* (22), eabf4741
12. De, A.; Bhunia, S.; Cai, Y.; Binyamin, T.; Etgar, L.; Ruhman, S. Spectator exciton effects in nanocrystals III: Unveiling the stimulated emission cross section in quantum confined CsPbBr3 nanocrystals. *Journal of the American Chemical Society* **2024,** *146* (29), 20241-20250
13. Cooney, R. R.; Sewall, S. L.; Sagar, D. M.; Kambhampati, P. Gain Control in Semiconductor Quantum Dots via State-Resolved Optical Pumping. *Phys. Rev. Lett.* **2009,** *102* (12), 4127404
14. Ryu, J.; Yeola, S.; Jonas, D. M. Generalized Einstein relations between absorption and emission spectra at thermodynamic equilibrium. *Proceedings of the National Academy of Sciences* **2024,** *121* (37), e2410280121
15. Cooney, R. R.; Sewall, S. L.; Sagar, D. M.; Kambhampati, P. State-resolved manipulations of optical gain in semiconductor quantum dots: Size universality, gain tailoring, and surface effects. *J. Chem. Phys.* **2009,** *131* (16), 13164706.
23

**Supplementary Information**

# Optical gain in colloidal quantum dots is limited by biexciton absorption, not biexciton recombination

Davide Zenatti[1] and Patanjali Kambhampati[1]*

[1]Department of Chemistry, McGill University, Montreal, QC H3A 0B8, Canada

*Correspondence to: pat.kambhampati@mcgill.ca

**Supplementary Information Preface**

This Supplementary Information provides the detailed theoretical framework underlying the microscopic description of optical gain presented in the main text. The purpose of this document is threefold: (i) to derive explicitly the physical constraints imposed by microscopic reversibility and the Einstein relations on optical gain; (ii) to formulate a minimal but complete exciton–biexciton–lattice model sufficient to describe gain thresholds and effective stimulated-emission cross sections in colloidal quantum dots; and (iii) to clarify how the two dimensionless control parameters introduced in the main text naturally emerge from this framework. The emphasis is on physical transparency and internal consistency rather than mathematical generality. The derivations presented here are self-contained and are intended to be accessible to advanced graduate students and researchers seeking a rigorous foundation for the results discussed in the main manuscript.

**Supplementary Table of Contents**

## 1. Scope and purpose of this Supplementary Information

The theory of optical gain in colloidal quantum dots has historically been formulated using phenomenological laser-level models that treat the quantum dot as an effective few-level system and invoke multiexciton decay dynamics, most notably Auger recombination, as the primary determinant of gain thresholds. While such models can rationalize certain qualitative trends, they do not enforce the fundamental constraints imposed by microscopic reversibility and therefore cannot provide a predictive or internally consistent description of gain across different material classes.

The purpose of this Supplementary Information is to present a minimal microscopic theory of optical gain that begins from constraints that must hold universally and proceeds systematically to quantities directly measured in experiment. In contrast to phenomenological approaches, the present framework does not assume a particular laser-level structure *a priori*, nor does it rely on population-counting arguments or rate-equation thresholds. Instead, optical gain is treated as a spectral balance problem governed by the relative placement and overlap of stimulated-emission and excited-state-absorption channels.

This Supplementary Information focuses exclusively on the theoretical foundations of the model introduced in the main text. Experimental methods, sample preparation, and spectroscopic techniques are discussed in the cited literature and are not repeated here. The central goal is to make explicit the physical assumptions, approximations, and derivations that lead to the expressions for the effective gain cross section, the gain-threshold condition, and the dimensionless control parameters η and χ used to construct the predictive phase diagram in Fig. 5 of the main manuscript.

Throughout this document, emphasis is placed on physical interpretation rather than algebraic generality. The theoretical treatment is intentionally restricted to the minimal level required to explain and predict observed gain phenomenology in colloidal quantum dots. More sophisticated treatments—including non-Markovian baths, higher-order multiexciton states, or strong-field coherent dynamics—are not required for the phenomena discussed here and are therefore outside the scope of this work.

## 2. Einstein relations and constraints on optical gain

Any physically consistent description of optical gain must begin from microscopic reversibility. For an optical transition between two many-body eigenstates $|i\rangle$ and $|f\rangle$ of energies $E_i$ and $E_f$, the probabilities for absorption and stimulated emission are not independent quantities but are linked by detailed balance. This connection is embodied in the Einstein relations, which impose exact constraints on optical cross sections independent of material composition, dimensionality, or disorder.

For a transition of angular frequency $\omega_{fi} = (E_f - E_i)/\hbar$, the absorption and stimulated-emission cross sections satisfy

$$\sigma_{\text{SE}}(\omega) = \frac{g_f}{g_i} \sigma_{\text{ABS}}(\omega), \qquad (S1)$$

where $g_i$ and $g_f$ are the degeneracies of the initial and final states, respectively. Equation (S1) is exact and follows directly from time-reversal symmetry and microscopic reversibility. Importantly, it implies that absorption and stimulated emission share the *same intrinsic spectral line shape*. Any difference between absorption and emission spectra must therefore arise from energetic displacement of the corresponding transitions, not from a change in the underlying line-shape function.

Equation (S1) has an immediate and often overlooked implication for optical gain. Because the intrinsic spectral form of stimulated emission is fixed by absorption, gain cannot be enhanced or suppressed by postulating an ad hoc emission spectrum that differs arbitrarily from absorption. Instead, any reduction in observed gain must originate from *additional absorptive channels* that overlap spectrally with stimulated emission and cancel it in the net response.

This distinction motivates a crucial separation between **bare** and **effective** stimulated-emission cross sections. The bare stimulated-emission cross section $\sigma_{\text{SE}}(\omega)$ is the Einstein-constrained quantity appearing in Eq. (S1) and is fixed once the absorption spectrum is known. Experiments, however, do not measure $\sigma_{\text{SE}}(\omega)$ directly. Instead, they measure the *net differential optical response* under excitation, which reflects the balance between stimulated emission and all excited-state absorption processes that are simultaneously active.

We therefore define the experimentally relevant effective gain cross section as

$$\sigma_{\text{eff}}(\omega) = \sigma_{\text{SE}}(\omega) - \sum_m \sigma_{\text{ESA}}^{(m)}(\omega), \qquad (S2)$$

where $\sigma_{\text{ESA}}^{(m)}(\omega)$ denotes excited-state absorption from the populated exciton manifold into higher-lying many-body states $m$. In colloidal quantum dots, the dominant contributions to the sum in Eq. (S2) arise from absorption into biexcitonic manifolds, as discussed in Sections 3 and 5.

Optical gain occurs only when $\sigma_{\text{eff}}(\omega) < 0$ over a finite spectral range. The condition for gain onset is therefore not a population-inversion criterion but a *spectral inequality* governed by the relative placement and overlap of stimulated-emission and excited-state-absorption channels. The gain threshold is determined by the condition that the spectrally integrated effective cross section vanishes,

$$\int d\omega\, \sigma_{\text{eff}}(\omega) = 0. \qquad (S3)$$

Equation (S3) defines the gain threshold independently of any assumptions about multiexciton lifetimes or decay rates. In particular, Auger recombination does not enter this condition and therefore cannot set the threshold for gain onset. Instead, Auger processes limit the *dynamics* and stability of gain once it is achieved, a distinction that is essential for interpreting experimental observations.

Finally, it is important to emphasize that the Einstein relations are never violated in quantum-confined systems. Apparent discrepancies between absorption-derived and experimentally measured stimulated-emission cross sections arise solely because experiments probe $\sigma_{\text{eff}}$, not $\sigma_{\text{SE}}$. Confusion between these two quantities has been a persistent source of misinterpretation in the quantum-dot gain literature. The framework developed here resolves this ambiguity by enforcing Einstein consistency at the microscopic level and explicitly accounting for all competing absorptive channels.

In the following sections, we identify the minimal electronic structure responsible for excited-state absorption in quantum dots and show how exciton–lattice coupling controls the spectral overlap entering Eqs. (S2) and (S3).

## 3. Minimal electronic structure: excitons and biexcitons

The microscopic origin of optical gain and loss in quantum dots cannot be understood within a single-transition picture. Even in the absence of lattice coupling, a quantum dot hosts a structured manifold of many-body electronic states that must be treated explicitly to identify all optically active channels.

The electronic ground state $|G\rangle$ is separated from the single-exciton sector by the exciton creation energy. Crucially, the exciton is not a single level but a *fine-structure manifold* $\{|X_\alpha\rangle\}$, where the index $\alpha$ labels states split by electron–hole exchange, spin–orbit coupling, and confinement anisotropy. These splittings are intrinsic to the electronic structure and exist even in perfectly rigid lattices. As a result, absorption and emission generally involve different members of the exciton manifold, giving rise to an intrinsic Stokes shift independent of lattice relaxation.

Upon optical excitation of an exciton, transitions into the two-exciton sector become allowed. The biexciton is likewise not a single level but a dense manifold $\{|XX_\beta\rangle\}$ of correlated two-electron–two-hole configurations. Coulomb interactions, exchange, and configuration interaction stabilize these states relative to two independent excitons and produce a broad distribution of biexciton energies. The biexciton manifold therefore constitutes a continuum of optically accessible final states rather than a discrete level.

A minimal electronic Hamiltonian sufficient to capture these features may be written schematically as

$$H_{\text{el}} = E_G \mid G\rangle\langle G \mid + \sum_\alpha E_{X_\alpha} \mid X_\alpha\rangle\langle X_\alpha \mid + \sum_\beta E_{XX_\beta} \mid XX_\beta\rangle\langle XX_\beta \mid. \qquad (S5)$$

Optical dipole operators couple $\mid G\rangle$ to $\mid X_\alpha\rangle$ and, once an exciton is present, couple $\mid X_\alpha\rangle$ to $\mid XX_\beta\rangle$. Consequently, *excited-state absorption (ESA) into biexcitonic states is unavoidable whenever the exciton population is nonzero*. This fact holds irrespective of biexciton lifetimes, excitation density, or pumping scheme.

The energetic placement of biexcitonic absorption relative to excitonic emission is conveniently characterized by the mean exciton-to-biexciton absorption detuning,

$$\Delta_{XX}^A \equiv \langle E_{XX_\beta} - E_{X_\alpha}\rangle, \qquad (S6)$$

where the average is taken over optically allowed transitions weighted by their oscillator strengths. This quantity governs where biexcitonic ESA appears relative to the emission band and therefore plays a central role in determining spectral overlap and gain cancellation.

It is important to emphasize that biexcitonic absorption competes directly with stimulated emission because both processes originate from the *same exciton population*. Increasing excitation density does not suppress ESA; instead, it enhances both emission and absorption channels proportionally. Optical gain is therefore controlled not by population inversion of a single transition but by the relative spectral placement and strength of these competing processes.

In the absence of lattice coupling, excitonic fine-structure splittings alone already enforce a minimal three-level structure, consisting of an absorptive exciton state and a lower-energy emissive exciton state. However, this three-level description remains lossy because ESA into biexcitonic manifolds overlaps spectrally with stimulated emission. As shown in subsequent sections, coupling to the lattice bath provides the additional degree of freedom required to modify this overlap and continuously transform the system toward effective four-level gain behavior.

In the next section, we introduce exciton–lattice coupling and show how it controls the spectral projection of the electronic structure into observable absorption and emission line shapes.

## 4. Exciton–lattice coupling and optical line-shape theory

The electronic structure introduced in Section 3 does not directly determine optical gain. What is measured experimentally are *optical spectra*, which reflect how electronic transitions are projected through coupling to the lattice environment. In colloidal quantum dots, excitons and biexcitons are embedded in a vibrational bath whose character ranges from narrow, crystalline phonon modes to broad, glassy or polaronic continua. This coupling plays a central role in shaping absorption, emission, and excited-state-absorption spectra.

To capture this physics at a minimal but sufficient level, we describe exciton–lattice coupling using a diagonal spin–boson Hamiltonian,

$$H = H_{\text{el}} + \sum_k \hbar \omega_k b_k^\dagger b_k + \sum_n |n\rangle\langle n| \sum_k g_{nk} (b_k^\dagger + b_k), \qquad \text{(S8)}$$

where $n \in \{X_\alpha, XX_\beta\}$ labels electronic states, $b_k^\dagger$ and $b_k$ are phonon creation and annihilation operators, and $g_{nk}$ describes state-dependent coupling to lattice modes of frequency $\omega_k$. This form assumes that lattice displacements depend on electronic occupation but neglects nonadiabatic electronic transitions induced by phonons, an approximation well justified for the optical processes considered here.

Electronic excitation displaces the equilibrium nuclear configuration, leading to lattice relaxation and spectral dressing of optical transitions. The strength and frequency dependence of this coupling are encoded in the phonon spectral density,

$$J_n(\Omega) = \sum_k |g_{nk}|^2 \delta(\Omega - \omega_k), \qquad \text{(S9)}$$

which fully determines the optical line shapes associated with transitions involving state $n$.

Within the independent-boson (or cumulant) formalism, the absorption or emission spectrum associated with an electronic transition of bare frequency $\omega_0$ may be written as

$$\sigma(\omega) \propto \text{Re} \int_0^\infty dt \, e^{i(\omega - \omega_0)t - g(t)}, \qquad (S10)$$

where the line-shape function $g(t)$ is given by

$$g(t) = \int_0^\infty d\Omega \, \frac{J(\Omega)}{\Omega^2} \left[ (1 - \cos \Omega t) \coth\left(\frac{\hbar\Omega}{2k_B T}\right) + i(\sin \Omega t - \Omega t) \right]. \qquad (S11)$$

This expression shows explicitly that optical spectra are not simple Gaussians but structured sums of phonon sidebands whose shape and width are determined by the bath spectral density. Absorption and emission share the same intrinsic line shape required by the Einstein relations but are energetically displaced due to lattice relaxation.

The energetic displacement between absorption and emission defines the reorganization energy,

$$\lambda = \int_0^\infty d\Omega \, \frac{J(\Omega)}{\Omega}, \qquad (S12)$$

which contributes to the observed Stokes shift. In quantum dots, this lattice-induced shift adds to the intrinsic Stokes shift arising from excitonic fine structure discussed in Section 3.

The relative importance of fine-structure splitting versus lattice-induced spectral broadening is captured by the dimensionless parameter

$$\eta \equiv \frac{\delta_{\text{FS}}}{W_{\text{ph}}},$$

where $\delta_{\text{FS}}$ is the characteristic exciton fine-structure splitting and $W_{\text{ph}}$ is the effective width of the phonon or polaronic progression. Large $\eta$ corresponds to crystalline systems with narrow phonon spectra, such as CdSe quantum dots, while small $\eta$ describes dynamically disordered or polaronically dressed lattices, such as lead-halide perovskite quantum dots.

As shown in the following section, lattice coupling not only broadens optical spectra but also controls the energetic placement and spectral overlap of biexcitonic absorption relative to excitonic emission. This provides the additional degree of freedom required to suppress excited-state absorption and enables the continuous crossover from lossy three-level behavior to effective four-level optical gain.

## 5. Excited-state absorption and definition of the biexciton proximity parameter χ

With the electronic structure and lattice coupling in place, we now identify the microscopic origin of the dominant loss channel competing with stimulated emission in quantum dots: excited-state absorption (ESA) from the single-exciton manifold into biexcitonic states. As emphasized in Sections 2 and 3, ESA is unavoidable whenever the exciton population is nonzero and therefore must be treated on equal footing with stimulated emission in any gain theory.

For a populated exciton state $|X_\alpha\rangle$, optical excitation accesses a manifold of biexcitonic states $|XX_\beta\rangle$. The corresponding ESA spectrum may be written as

$$\sigma_{\text{ESA}}^{XX}(\omega) = \sum_{\alpha,\beta} P_{X_\alpha} \ |\mu_{X_\alpha \to XX_\beta}|^2 \ L_{XX_\beta, X_\alpha}(\omega), \qquad \text{(S13)}$$

where $P_{X_\alpha}$ is the population of exciton state $\alpha$, $\mu_{X_\alpha \to XX_\beta}$ is the optical transition dipole moment, and $L_{XX_\beta, X_\alpha}(\omega)$ is the Einstein-consistent line shape for the exciton-to-biexciton transition, including lattice dressing as described in Section 4.

The energetic placement of this absorptive manifold relative to excitonic emission is the key determinant of spectral overlap and gain cancellation. To characterize this placement in a dimensionless and material-independent way, we define the biexciton proximity parameter

$$\chi \equiv \frac{\Delta_{XX}^{A}}{\Delta_{SS}}, \qquad \text{(S14)}$$

where $\Delta_{XX}^A$ is the mean exciton-to-biexciton absorption energy defined in Eq. (S6), and $\Delta_{SS}$ is the total excitonic Stokes shift, including contributions from both fine structure and lattice relaxation.

The parameter $\chi$ quantifies how deeply absorptive biexcitonic transitions encroach into the emission band. When $\chi \approx 1$, biexcitonic ESA is spectrally correlated with excitonic emission, leading to strong cancellation between stimulated emission and absorption. When $\chi < 1$, biexcitonic absorption is displaced away from emission, reducing spectral overlap and suppressing loss.

Importantly, $\chi$ is not an abstract fitting parameter but encodes concrete physical properties of the quantum dot. Biexciton stabilization energies depend on Coulomb correlations, dielectric screening, and carrier localization, while the Stokes shift reflects both excitonic fine structure and lattice reorganization. Surface trapping provides an additional pathway for reducing $\chi$: surface-localized excitons form more strongly stabilized biexcitonic configurations, shifting ESA deeper into the emission band and increasing loss. This provides a natural microscopic explanation for the observed correlation between surface passivation, photoluminescence quantum yield, and gain thresholds discussed in the main text.

Together with the lattice-coupling parameter $\eta$, the biexciton proximity parameter $\chi$ fully specifies the spectral geometry governing optical gain. In the next section, we combine these parameters to construct the effective gain cross section and derive the gain-threshold condition used to generate the phase diagram in Fig. 5 of the main manuscript.

## 6. Effective gain cross section and gain-threshold condition

Optical gain in quantum dots emerges from the competition between stimulated emission from the exciton manifold and excited-state absorption into biexcitonic manifolds. As shown in Section 2, experiments probe the *effective* gain cross section rather than the bare Einstein-constrained stimulated-emission cross section. We now make this distinction explicit and derive the gain-threshold condition used throughout the main text.

For a given exciton population, the effective gain cross section is defined as

$$\sigma_{\text{eff}}(\omega) = \sigma_{\text{SE}}(\omega) - \sigma_{\text{ESA}}^{XX}(\omega), \qquad \text{(S16)}$$

where $\sigma_{\text{SE}}(\omega)$ is the bare stimulated-emission cross section fixed by the absorption spectrum through the Einstein relations, and $\sigma_{\text{ESA}}^{XX}(\omega)$ is the biexciton-mediated excited-state absorption defined in Eq. (S13). Both terms scale linearly with exciton population and therefore cannot be separated by population arguments alone.

Optical gain occurs when $\sigma_{\text{eff}}(\omega)$ becomes negative over a finite spectral interval. The onset of gain is determined by the condition that the *spectrally integrated* effective cross section vanishes,

$$\int d\omega \, \sigma_{\text{eff}}(\omega) = 0. \qquad \text{(S17)}$$

Equation (S17) defines the gain threshold in a manner that is independent of excitation dynamics, pumping geometry, or cavity feedback. It depends solely on the relative spectral placement and overlap of emission and absorption channels and therefore constitutes a purely spectral criterion for gain onset.

Crucially, multiexciton lifetimes do not appear in Eq. (S17). Auger recombination limits the *duration* over which gain can be sustained once it is achieved, but it does not determine whether gain is achieved in the first place. This distinction explains why gain thresholds extracted from state-resolved optical pumping experiments do not scale with biexciton lifetimes or with inverse quantum-dot volume, contrary to expectations based on phenomenological two-level models.

The solution of Eq. (S17) yields the average exciton number at threshold, $\langle N_{\text{th}} \rangle$, which depends only on the two dimensionless parameters introduced above: the lattice-coupling parameter $\eta$ and the biexciton proximity parameter $\chi$,

$$\langle N_{\text{th}} \rangle = F(\eta, \chi), \qquad (S18)$$

where the function $F$ is determined by the Einstein-consistent line shapes of stimulated emission and biexcitonic absorption. In the crystalline limit ($\eta \gg 1$) and for $\chi \approx 1$, emission and absorption remain spectrally correlated, enforcing strong cancellation and pinning $\langle N_{\text{th}} \rangle \approx 1$. As either $\eta$ or $\chi$ is reduced, spectral overlap decreases and the gain threshold is correspondingly lowered.

An analogous expression governs the experimentally measured stimulated-emission cross section. Integrating Eq. (S16) over frequency yields

$$\sigma_{\text{eff}} = \int d\omega \, \sigma_{\text{eff}}(\omega), \qquad (S19)$$

which is reduced relative to the Einstein-limited value whenever biexcitonic absorption overlaps spectrally with emission. The same parameters $\eta$ and $\chi$ therefore control both the gain threshold and the effective stimulated-emission cross section, providing a unified microscopic explanation for two long-standing and previously disconnected observations in quantum-dot gain experiments.

In the final section, we summarize the controlled limits of this theory and clarify how crystalline CdSe and lattice-dressed quantum dots occupy distinct regions of the $(\eta, \chi)$ phase space defined in the main text.

## 7. Controlled limits: crystalline CdSe and lattice-dressed quantum dots

The microscopic framework developed in this Supplementary Information admits two experimentally relevant and physically transparent limits, which correspond to distinct regions of the $(\eta, \chi)$ parameter space introduced in the main text. These limits clarify why different classes of colloidal quantum dots exhibit qualitatively different gain behavior despite sharing similar electronic architectures.

In crystalline covalent quantum dots such as CdSe, exciton–lattice coupling is relatively weak and dominated by narrow phonon spectra. As a result, the lattice-coupling parameter is large ($\eta \gg 1$), reflecting the fact that excitonic fine-structure splittings are comparable to or larger than the width of the phonon progression. In this regime, absorption and emission spectra remain strongly correlated, and biexcitonic absorption tracks excitonic emission closely when $\chi \approx 1$. The resulting spectral overlap enforces substantial cancellation between stimulated emission and excited-state absorption, pinning the gain threshold near one exciton per dot and suppressing the effective stimulated-emission cross section. This behavior is robust against changes in dot size and cannot be overcome by modifying multiexciton lifetimes, consistent with decades of experimental observations in CdSe quantum dots.

In contrast, in dynamically disordered or polaronically dressed lattices—such as lead-halide perovskite quantum dots—exciton–lattice coupling is strong and broadband, yielding small values of $\eta$. In this limit, lattice dressing produces a continuous displacement between absorption and emission surfaces, decorrelating biexcitonic absorption from excitonic emission even when biexciton stabilization remains finite. When combined with reduced biexciton proximity ($\chi < 1$), this decorrelation suppresses spectral overlap and drives a continuous crossover toward effective four-level gain behavior. Both the gain threshold and the effective stimulated-emission cross section are thereby enhanced without invoking changes in carrier lifetimes or excitation dynamics.

These two limits are unified by the same microscopic theory and differ only in the relative magnitudes of $\eta$ and $\chi$. The resulting phase diagram provides a direct materials-design principle: optimal gain performance is achieved by simultaneously minimizing biexciton-mediated spectral loss (reducing $\chi$) and exploiting lattice disorder or polaron formation to broaden the phonon spectral density (reducing $\eta$).

## 8. Summary of theoretical results and connection to the main text

This Supplementary Information has developed a minimal, Einstein-consistent microscopic theory of optical gain in colloidal quantum dots. By treating excitons, biexcitons, and lattice coupling on equal footing, the theory identifies excited-state absorption into biexcitonic manifolds as the dominant loss channel limiting gain performance and shows that gain onset is governed by spectral overlap rather than by population inversion or multiexciton lifetimes. Two dimensionless parameters—the lattice-coupling parameter $\eta$ and the biexciton proximity parameter $\chi$—emerge naturally from the theory and fully specify the gain threshold and effective stimulated-emission cross section. These results provide the formal foundation for the predictive phase diagram presented in Fig. 5 of the main manuscript and establish a unified framework for understanding and engineering optical gain across all major classes of colloidal quantum dots.